Geometric and physical considerations for realistic protein models.

Isaac A. Hubner and Eugene I. Shakhnovich[*]

Department of Chemistry and Chemical Biology
Harvard University
12 Oxford Street
Cambridge, MA 02138

[*]corresponding author
tel: 617-495-4130
fax: 617-384-9228
email: eugene@belok.harvard.edu

**Abstract**

Protein structure is generally conceptualized as the global arrangement or of smaller, local motifs of helices, sheets, and loops. These regular, recurring secondary structural elements have well-understood and standardized definitions in terms of amino acid backbone geometry and the manner in which hydrogen bonding requirements are satisfied. Recently, "tube" models have been proposed to explain protein secondary structure in terms of the geometrically optimal packing of a featureless cylinder. However, atomically detailed simulations demonstrate that such packing considerations alone are insufficient for defining secondary structure; both excluded volume and hydrogen bonding must be explicitly modeled for helix formation. These results have fundamental implications for the construction and interpretation of realistic and meaningful biomacromolecular models.

# 1. Introduction

Ask any physicist, chemist, or biologist about protein structure, and they will likely begin with a description of α helices and β sheets. These regular secondary structural elements and their tertiary arrangement form the basis for understanding the beautiful recurring patters of protein topology, which separate proteins ("evolved" or "designed" heteropolymer) from random heteropolymers in that proteins achieve a distinct native conformation that has a sizable energy "gap" from all other states[1]. This state is characterized not only by secondary structural motifs, but a unique compact topology in which the interior of the protein has the packing density of a molecular crystal[2]. Despite serving an extensive range of physiological functions and varying widely in primary sequence (which allows for virtually innumerable combinations of the twenty distinct amino acids) proteins have well described local geometric regularities. The geometric restraints resulting from chemical connectivity and excluded volume, which were first systematically explored through the use of simple models[3], are quantified by allowable regions of $\phi$ and $\psi$ backbone angles (Figure 1).

Questions regarding the requirements of a protein model realistic enough to be biologically and physically meaningful seem especially timely in light of recent papers[4-7], which have suggested that the form and formation of protein secondary structure is a geometrical consequence of the collapse of "tubes". This class of models represents the amino acid chain as a flexible, self-avoiding cylinder of finite radius with an interaction potential representing hydrophobic forces. One review, which explicitly describes a protein as a "tube", proposes that the details of the amino acids are not important in determining protein fold; implying that protein structure, specifically α helices and β sheets and extending to compact conformations, are explainable as "ground-state structures associated with the marginally compact phase of a short tube"[8]. In each case, the observation of helices is a geometric result of efficient tube packing upon energy minimization while maintaining excluded volume, challenging the view that hydrogen bonding plays an important role in protein structure.

The basic requirements of a biophysically relevant protein model are 1) a representation of chain connectivity and excluded volume, 2) recurring local structural motifs that assemble into a unique, compact global structure, and 3) an energy function that places this "native" conformation as the global energy minimum. The tube model predicts that geometrical and topological factors alone, without inclusion of more chemically detailed hydrogen bonding interactions, determine global features of protein folds such as protein-like secondary structure[4, 8]. Therefore, if tube models have implications for real proteins, one would expect similar formation, upon collapse, of helices and secondary structure motifs in a model that accurately represented the geometric and topological properties of amino acid chain in terms of excluded volume and torsional degrees of freedom (as opposed to a featureless tube), but is devoid of explicit hydrogen bonding. Here, we explore the requirements of a spatially and geometrically realistic polymer chain model to achieve protein-like behavior. Simulations demonstrate that simple collapse of geometrically realistic polymer chains is insufficient to produce protein-like secondary structure. Rather, an explicit representation of hydrogen bonding, in addition to a potential to drive polymer collapse are necessary. In building models that are descriptive and predictive of protein structure and behavior, the underlying physical motivation must be carefully considered.

## 2. Model and Results

In a spatially and geometrically realistic protein model, does a hydrophobic-driven collapse produce the helical structures suggested by the above simpler "tube" models? By spatially and geometrically realistic, we mean that protein coordinates representing all atoms are represented as impenetrable hard spheres of physical radii (excluded volume) and connected by bonds with free rotation about $\phi$, $\psi$, and $\chi$ angles (Figure 1) while maintaining the planarity of the peptide bond. Simulations with this spatial and geometric representation, propagated *via* Monte Carlo dynamics have previously been described in detail[9, 10]. In the simulations presented here, we follow the protocol presented in [11], using a knowledge-based atomic interaction potential that has previously been applied to folding SH3 domains[12] and has been shown to predict the folding and structure of six different small helical proteins[11]. Using this model we compared the results of protein folding simulations with and without explicit hydrogen bonding (an attractive directional interaction between backbone nitrogen and carbonyl groups, also detailed in [11]) to explore the applicability of the conclusions from the tube model to more realistic protein models. If polymer "thickness" and hydrophobicity are sufficient for folding, or at least forming secondary structure, as tube models claim, one would expect to observe at least some helix or sheet geometries, in both simulations.

While this protein model has previously been described in detail (we refer the reader to [11] for an in depth explanation), the motivation for the energy function warrants brief discussion. Behavior of a polymer chain can be described in terms of the generalized microscopic Hamiltonian:

$$H(\{r_i\},\{\sigma_i\}) = \sum_{i<j} B(\sigma_i,\sigma_j) U(r_i - r_j) \quad (1)$$

where the energy is defined by the position ($r_i$) and identity ($\sigma_i$) of each atom based on chemical identity (through an interaction matrix $B$) and pairwise distances (through some potential $U$). In considering the form of the interaction potential, a full quantum-mechanical treatment is impossible but, also (fortunately) unnecessary as we are not modeling the formation or breaking of covalent bonds. The energy of a protein conformation can be estimated by the sum of the hydrogen bonding and van der Waals interactions (if disulfide bonds and salt bridges are not present, as in the proteins studied here, ionic and covalent interactions in folding can be, for all practical purposes, excluded). Van der Waals interactions are non-directional and on the order of ~0.1-1 kcal/mol while hydrogen bonds are highly dependent upon geometry (in the case of a sp$^2$ hybridized acceptor such as a carbonyl oxygen) and are on the order of ~1-5 kcal/mol[13]. In this respect, it is reasonable to represent the Hamiltonian as two separate sums of hydrogen bonding energies over hydrogen bonding atoms (backbone carbonyl oxygen and amide hydrogen) and van der Waals interactions over the remaining sidechain atoms given by:

$$H_{Total} = H_{HB} + H_{vdW} = \sum_{i<j} B_{HB} U_{HB} + \sum_{i<j} B_{vdW} U_{vdW} \quad (2)$$

Where $B_{HB}$ and $B_{vdW}$ are the interaction matrices for hydrogen bonding, and atom-atom interaction, respectively. It is also noteworthy that, due to differing spatial and geometric dependencies, the potentials for these functions ($U_{HB}$ and $U_{vdW}$) are of different forms.

For this test, we choose two extensively studied small helix bundle proteins: 1BDD and 1ENH. Folding from random coil with the full model produces structures ~3Å from the native state with helices and well-packed hydrophobic cores (Table 1 and Figure 2). However, when the same model is applied with only the hydrophobic interaction term and without hydrogen bonding, no helices are observed. Secondary structure content was objectively determined according to the standardized definition of DSSP[14]. Structures "folded" without explicit hydrogen bonding have on average ~1% of residues in helical geometries as compared to the ~50% helical geometry of the native-like structures folded with hydrogen bonding. Measures of collapse such as radius of gyration ($R$g) and number of contacts ($N$) show the conformations are optimally (native-like) compact. Despite efficient packing of a realistic peptide chain driven by hydrophobic collapse, without an explicit representation of hydrogen bonding, helix geometries do not form.

Since protein folding is an ensemble process, we compare the results of 200 independent runs for each model. Clustering minimum energy conformations from each of run by structural similarity ($C_\alpha$ RMSD)[11] reveals an interesting behavior. While simulations conducted with the full potential segregated largely into a dense cluster of similar (native-like) conformations, simulations without hydrogen bonding produced many small, disparate (non-native) clusters. Simulations without hydrogen bonding resulted in conformations that resemble neither the native state nor each other. A spatially and geometrically realistic protein chain modeled without hydrogen bonding fails another important test of protein behavior: the energetic ground state is not structurally unique or well defined. It appears that hydrogen bonding not only helps define local secondary structure, but also is essential to the requirement that a protein native state be topologically distinct and energetically separated from all other states. Hydrogen bonding functions as a structural restraint, decreasing the number of low energy misfolded states ("decoys'') with which native conformation has to compete.

## 3. Comments and Conclusions

Clearly, compaction of a realistic protein chain model without consideration of hydrogen bonding does not necessarily result in helical geometries. Packing is an undeniably important consideration in protein folding. A protein's interior is as tightly packed as a molecular crystal[2], maximizing favorable van der Waals interactions. The importance of excluded volume in peptide geometry is a well-established fact. G. N. Ramachandran first formalized the general effect on chain geometry in the now ubiquitous "Ramachandran plots", which relate sterically allowed regions of ϕ and ψ with secondary structural[3]. In contrast to tube models, Ramachandran plots show that sterically allowed regions correspond predominantly to regions of well-defined secondary structure in terms of the backbone geometry of individual amino acids, not that excluded volume defines secondary structure. Excluded volume and packing of a "tube" are insufficient to explain or predict protein secondary structure or folding. Without the specific geometric requirements imposed by hydrogen bonding, polymer collapse does not produce helices or sheets. It was the keen consideration of this detail that led Linus Pauling to the first predictive model of protein secondary structure geometry[15].

Tubes may be parameterized (r and t in [5]) and compacted to resemble a protein's helix pitch and rise, but the similarity is skin deep. One study explained β sheets by a change in the relative sizes of the solvent and tube[5], a phenomena that is not observed in nature. While it is undeniable that simple models have played an important role in understanding protein folding[16], it is dangerous to generalize to explain phenomena that are beyond the model's capacity. As simulations of protein folding, structure prediction, and biomacromolecular systems in general reach ever higher levels of sophistication and complexity, it is of vital importance that the theoretical concepts on which they are based and motivated are physically sound.

Ultimately, sidechains should be considered in some way as their hydrophobic collapse provides the energetic driving force for folding[13]. Sidechains are also important as it has long been known that different residues have propensities for particular types of secondary structure[17]. Even in the approximation of ignoring sidechains, both hydrophobic packing and the inevitable burial of polar groups (backbone amides and carbonyls), which are nearly always hydrogen bonded[2], must be considered in understanding secondary structure. Proteins fold in an aqueous environment so hydrogen bonds do not contribute energetically to the folded state relative to the unfolded state. However, without intra-protein hydrogen bonding, folding would face an insurmountable energetic penalty from the loss of hydrogen bonds to water upon collapse. The requirement of backbone hydrogen bonding is serious geometric constraint, which along with the steric restraints of the peptide chain defines secondary structure. It is certainly serendipitous that meeting this geometric constraint coincides with a nearly optimal geometric packing of atoms, but would nature have done it any other way?

**Acknowledgments.** We are grateful to A. Grosberg for useful discussions. This work was supported by a Howard Hughes Medical Institute fellowship (to IAH) and the NIH.

**Table 1.** Comparison of two proteins folded with and without explicit hydrogen bonding. "Native" refers to the RCSB PDB structure, whereas "Model" and "No HB" are the protein model (described in [11]) with and without explicit hydrogen bonding. "% helical" is the number of i → i+4 backbone hydrogen bond geometries per 100 residues, calculated by DSSP[14]. "*Rg*" is the radius of gyration and "*N*" is the number of contacts. Values are averaged over 200 independent structures. All conformations are of native compactness (as measured by *Rg*) and, in fact, the structures modeled without hydrogen bonding have a higher contact density (*N*, indicative of sidechain packing).

|      |        | % helical | *Rg*  | *N*  |
|------|--------|-----------|-------|------|
| 1BDD | Native | 65        | 9.68  | 398  |
|      | Model  | 45        | 10.07 | 525  |
|      | No HB  | 1         | 10.37 | 609  |
| 1ENH | Native | 56        | 10.73 | 236  |
|      | Model  | 47        | 11.39 | 395  |
|      | No HB  | 1         | 11.54 | 471  |

**Figure 1.** A protein chain fragment illustrating the torsional degrees of freedom of an individual amino acid (ϕ, ψ, and χ angles) and its chemical bonding to neighboring amino acids thorough peptide bonds. The CO-NH linkage has approximately 40% double bond character through resonance with the carbonyl to form a Schiff base, rendering it effectively planar. The R group may represent any amino acid sidechain.

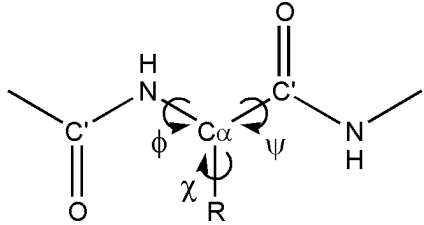

**Figure 2.** Protein models from the PDB and representatives from simulation. Model simulations fold to native conformations while simulations without hydrogen bonding collapse without helices. Excluded volume and an attractive potential ensure a protein-like hydrophobic core and sidechain packing. However, representation of hydrogen bonding interactions is essential for formation of secondary structure.

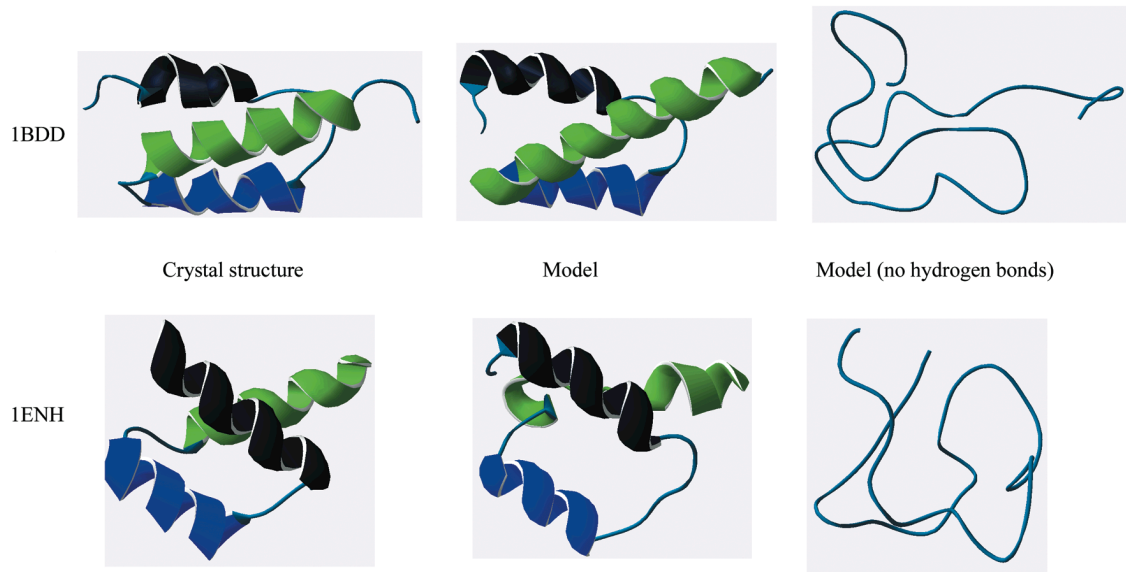